\documentclass[aps,prd,preprint,superscriptaddress,amsmath,amssymb,showpacs,article]{revtex4-1}
\usepackage{dcolumn}
\usepackage{graphicx}
\usepackage{float}
\usepackage{physics}
\usepackage[colorlinks=true,allcolors=blue]{hyperref}
\usepackage{float}
\usepackage{graphicx}

\begin{document}

\title{String-breaking of the triply heavy baryon at finite temperature and chemical potential}

\author{Jia-Jie Jiang}
\affiliation{School of Nuclear Science and Technology, University of South China, Hengyang 421001, China}

\author{Xun Chen}
\email{chenxunhep@qq.com}
\affiliation{School of Nuclear Science and Technology, University of South China, Hengyang 421001, China}

\author{Jiajia Qin}
\email{jiajiaqin@usc.edu.cn}
\affiliation{School of Nuclear Science and Technology, University of South China, Hengyang 421001, China}
\date{\today}

\author{Miguel Angel Martin Contreras}
\email{miguelangel.martin@usc.edu.cn}
\affiliation{School of Nuclear Science and Technology, University of South China, Hengyang 421001, China}
\date{\today}

\begin{abstract}
In this work, we use gauge/gravity duality to study potential energy and string-breaking of triply heavy baryons at finite temperature and chemical potential. Two different possible configurations of triply heavy baryon are considered. The effect of temperature and chemical potential on string-breaking distance is investigated. With increasing temperature/chemical potential, the sting-breaking distance decreases and then increases in symmetric collinear geometry while steadily increasing in equilateral triangle geometry. Our study provides insights into triply heavy baryon configuration and string-breaking behavior.
\end{abstract}

\maketitle

\section{Introduction}\label{int}
By studying the triply heavy baryon potential model, we can deepen our understanding of the formation of the quark-gluon plasma (QGP), which lacks adequate explanation in the framework of the current theories of heavy quark structure and properties\cite{Flynn:2011gf, Andreev:2015riv}.
AdS/CFT is a useful means to deal with QCD problems. The results can provide important information about the QCD properties of the non-perturbative energy region\cite{Maldacena:1997re, Gubser:1998bc, Witten:1998qj}.

In the SU(3) quenched lattice QCD, the three-quark potential is found to be well reproduced by
\begin{equation}
V_{3 Q}\left(\boldsymbol{r}_1, \boldsymbol{r}_2, \mathbf{r}_3\right)=\sigma_{3 Q} L_{\min }-\sum_{i<j} \frac{A_{3 Q}}{\left|\boldsymbol{r}_i-\boldsymbol{r}_j\right|}+C_{3 Q} .
\end{equation}
Here $\boldsymbol{r}_1, \boldsymbol{r}_2$, and $\boldsymbol{r}_3$ are the positions of the three quarks, and $L_{\min }$ is the minimum flux-tube length connecting the three quarks.
The strength of quark confinement is controlled by the string tension $\sigma_{3 \mathrm{Q}}$ of the flux tube.
In the hadron model, baryons are composed of three quarks, which are linked together by a Y-shaped configuration \cite{Artru:1974zn, Rossi:1977cy, Isgur:1984bm, Rossi:2016szw, Maldacena:1998im}.
Oleg Andreev has proposed effective multi-quark potential models from holography at vanishing temperature and chemical potential\cite{Andreev:2015riv, Andreev:2015iaa, Andreev:2019cbc, Andreev:2020xor, Andreev:2021bfg, Andreev:2021eyj, Andreev:2022cax}, we recently studied the triply heavy baryon at finite temperature and chemical potential from holography\cite{Jiang:2022zbt}.

The fragmentation methods of the three-quark $QQQ$ are as follows\cite{Andreev:2021bfg}
\begin{gather}
\begin{aligned}
\begin{matrix}
&\nearrow QQq+Q\bar{q}\\
\\QQQ& \to Qq+2Q\bar{q}\bar{q}\\
\\ &\searrow qqq+3Q\bar{q}.
\end{matrix}
\end{aligned}
\end{gather}
According to lattice calculations in SU(3) gauge theory, the potential energy between three heavy quarks increases linearly with distance as they are pulled apart\cite{Kadam:2022ipf}.
The potential energy becomes saturated when the distance between quarks reaches a certain length, causing the string connecting the three quarks to break.
As a result, new quark states and quark-antiquark pairs are generated\cite{Baker:2019gsi}.
In string models of hadrons, such a phenomenon is interpreted as string-breaking\cite{Artru:1974zn, Rossi:1977cy, Isgur:1984bm}.
This particular distance sets a characteristic scale called the string-breaking distance\cite{Drummond:1998ar, Bulava:2019iut}. This scale is obtained from the following formula
\begin{equation}
E_{\mathrm{Q} \overline{\mathrm{Q}}}\left(\boldsymbol{\ell}_{\mathrm{Q} \overline{\mathrm{Q}}}\right)=2 E_{\mathrm{Q} \overline{\mathrm{q}}} \text {. }
\end{equation}
We apply the same method to determine the string-breaking length in the three-quark state.
Three quarks exhibit different potential energies at the same chord distance depending on finite temperatures and chemical potentials.
If we assume that hadrons are non-interacting, the total energy is the sum of the energies of individual hadrons \cite{Andreev:2021bfg}.
When we observe that quark models of different states possess equal energy and their linear potentials are identical, it is inferred that they have the potential to undergo string-breaking.

The study of the triple heavy quark potential and its string breaking has been one of the key issues in quantum chromodynamics\cite{Bali:2000gf}.
QCD string fracture research has a long history\cite{Bali:2000vr, UKQCD:1998zbe, Michael:1999nq, Pennanen:2000yk, Schilling:1999mv}. The lattice calculations and AdS/CFT approaches have been studied accordingly\cite{Campbell:1985kp, Michael:1991nc, Szasz:2008qk}.
The existing literature primarily focuses on the stability and decay mode of triple heavy quarks under zero-temperature chemical potential conditions\cite{Flynn:2011gf,  Baker:2019gsi, Bali:2005fu}, while the effect of the temperature chemical potential is less considered\cite{Jiang:2022zbt, Antonov:2023uck}. In order to gain a deeper understanding of the configuration behavior of triple heavy quarks, we have systematically investigated for the first time the effect of the temperature chemical potential on the string-breaking distance of triple heavy quarks using holographic theory. Studying the string-breaking process helps to reveal the breaking mechanism of triple-heavy quarks.

The article is organized as follows. In Sec.\textmd{\ref{Model}}, we will provide the potential energy calculation formulas for $QQQ$, $QQq$,$Q\bar{q}$,$Qqq$ and $qqq$, respectively. In Sec.\textmd{\ref{Numerical results}}, we will draw potential energy diagrams for various hadrons and calculate the string-breaking results.
In Sec.\textmd{\ref{Summary and conclusion}}, we will summarize the results of our work.

\section{Model}\label{Model}
We choose the deformed AdS-RN metric. It is not a self-consistent solution to the Einstein equation. However, it gains some insights into problems without predictions from phenomenology and lattice QCD.
The background is of the form\cite{Andreev:2015riv,Andreev:2006nw}
\begin{equation}\label{metric0}
\begin{gathered}
d S^{2}=e^{s r^{2}} \frac{R^{2}}{r^{2}}\left[f(r) d t^{2}+d \vec{x}^{2}+f^{-1}(r) d r^{2}\right], \\
f(r)=1-\left(\frac{1}{r_{h}^{4}}+q^{2} r_{h}^{2}\right) r^{4}+q^{2} r^{6}.
\end{gathered}
\end{equation}
Here, $q$ is the black hole charge,  and $r_{h}$ is the position of the black hole horizon.
The Hawking temperature of the black hole is defined as
\begin{equation}
T=\frac{1}{4 \pi}\left|\frac{d f}{d r}\right|_{r=r_{h}}=\frac{1}{\pi r_{h}}\left(1-\frac{1}{2} Q^{2}\right),
\end{equation}
\noindent where $Q=q r_h^3$ and $0 \leq Q \leq \sqrt{2}$\cite{Herzog:2006ra}.
It is observed that, on dimensional grounds, the low $r$ behavior of the bulk gauge field $A_0(r)$ is\cite{Colangelo:2010pe}
\begin{equation}
A_0(r) = \mu - \eta r^2,
\end{equation}
$\eta = \kappa q$  with $\kappa$ a dimensionless parameter.
Together with the condition that $A_0$ vanishes at the horizon $A_0(r_h) = 0$  and $A_{0}(0)=\mu$, we can determine a relation between $\mu$ and the black-hole charge $q$:
\begin{equation}
\mu= q \kappa r_h^2 = \kappa \frac{Q}{r_{h}}.
\end{equation}
We fix the parameter $\kappa$ to 1 in this paper.
The AdS/CFT dictionary defines $\mu$ as the baryon number chemical potential.
Thus, we can get
\begin{equation}
\begin{gathered}
f(r)=1-\left(\frac{1}{r_{h}^{4}}+\frac{\mu^{2}}{r_{h}^{2}}\right) r^{4}+\frac{\mu^{2}}{r_{h}^{4}} r^{6}, \\
T=\frac{1}{\pi r_{h}}\left(1-\frac{1}{2} \mu^{2} r_{h}^{2}\right), \\
A_{0}(r)=\mu-\mu \frac{r^{2}}{r_{h}^{2}}.
\end{gathered}
\end{equation}

The first is the fundamental string governed by the Nambu-Goto action.
By using the static gauge $t=\tau, \sigma=x$, the Nambu-Goto action of the U-shape string is \cite{Zhang:2015faa}
\begin{equation}\label{sng}
S_{NG}=\frac{1}{2 \pi \alpha^{\prime}} \int d \tau d \sigma \sqrt{\operatorname{det} g_{\alpha \beta}},
\end{equation}
with
\begin{equation}
g_{\alpha \beta}=G_{\mu v} \frac{\partial x^{\mu}}{\partial \sigma^{\alpha}} \frac{\partial x^{v}}{\partial \sigma^{\beta}},
\end{equation}
where $X^{\mu}$ and $G_{\mu v}$ are the target space coordinates and the metric, respectively, and $\sigma^{\alpha}$ with $\sigma = 0,\,1$ parameterize the world-sheet.

The baryon vertex is given by\cite{Chen:2021bkc,Andreev:2020pqy}
\begin{equation}
S_{\mathrm{vert}}=\tau_{v} \int \mathrm{d} t \frac{\mathrm{e}^{-2 s r^{2}}}{r} \sqrt{f(r)},
\end{equation}
where $\tau_{v}$ is a dimensionless parameter defined by $\tau_{v}=$ $\mathcal{T}_{5} R \operatorname{vol}(\boldsymbol{X})$, and $\operatorname{vol}(\boldsymbol{X})$ is a volume of $\boldsymbol{X}$.
We consider the light quarks at string endpoints as tachyon fields. Thus, we add to the world-sheet action $S_{\mathrm{q}}=\int d \tau e \mathrm{~T}$, which is the usual sigma-model action for strings propagating in a tachyon background\cite{Andreev:2020pqy}.
The action written in the static gauge is
\begin{equation}
S_{\mathrm{q}}=\mathrm{m} \int d t \frac{\mathrm{e}^{\frac{\mathrm{s}}{2} r^2}}{r} \sqrt{f(r)}.
\end{equation}
where $\mathrm{m}=R\mathrm{~T}_0$.
We immediately recognize it as the action of a point particle of mass $\rm T_0$ at rest \cite{Bars:1977ud}.
In the presence of a background gauge field, the string endpoints with attached quarks couple to it.
So the world-sheet action includes boundary terms given by 
\begin{equation}
S_{A}=\mp \frac{1}{3} \int d t A_0,
\end{equation}
The minus plus sign corresponds to a quark and an antiquark, respectively.
We will choose model parameters as follows: $\mathrm{g}=\frac{R^2}{2 \pi \alpha^{\prime}},\, \mathrm{k}=\frac{\tau_v}{3\mathrm{~g}}$, and $\mathrm{n}=\frac{\mathrm{m}}{\mathrm{g}}$.

\subsection{QQQ model}
The total action of three-quark potential is the sum of the three Nambu-Goto actions plus the vertex action and the boundary action terms
\begin{equation}
S=\sum_{i=1}^{3} S_{N G}^{(i)}+S_{\mathrm{vert}}+3 S_A|_{r=0}.
\end{equation}
Thus, the action can now be written as
\begin{equation}
\begin{gathered}
S=\frac{3 \mathrm{g}}{\mathrm{T}} \int d x \frac{e^{s r^{2}}}{r^{2}} \sqrt{f(r)+\left(\partial_{x} r\right)^{2}}+\frac{3 \mathrm{k} \mathrm{g}}{\mathrm{T}} \frac{e^{-2 s r_{v}^{2}}}{r_{v}} \sqrt{f(r_{v})}-\frac{A_{0}(0)}{\mathrm{T}},
\end{gathered}
\end{equation}
with $\mathrm{T}=\int_0^\mathrm{T} d t$.
The Lagrangian is written as
\begin{equation}
\mathcal{L}=\frac{e^{s r^{2}}}{r^{2}} \sqrt{f(r)+\left(\partial_{x} r\right)^{2}},
\end{equation}
altogether, with the constraint
\begin{equation}
\mathcal{L}-r^{\prime} \frac{\partial\mathcal{L}}{\partial r^{\prime}}=\frac{\frac{e^{s r^{2}}}{r^{2}} f(r)}{\sqrt{f(r)+\left(\partial_{x} r\right)^{2}}}=\rm constant.
\end{equation}
At the points $r_0$ and $r_v$, we have
\begin{equation}
\begin{aligned}
&\frac{\frac{e^{s r^{2}}}{r^{2}} f(r)}{\sqrt{f(r)+\left(\partial_{x} r\right)^{2}}}=\frac{e^{s r_{0}^{2}}}{r_{0}^{2}} \sqrt{f\left(r_{0}\right)} , \\
&\frac{\frac{e^{s r_{v}^{2}}}{r_{v}^{2}} f\left(r_{v}\right)}{\sqrt{f\left(r_{v}\right)+\tan ^{2} \alpha}}=\frac{e^{s r_{0}^{2}}}{r_{0}^{2}} \sqrt{f\left(r_{0}\right)}.
\end{aligned}
\end{equation}
The above equations can determine the relation between $r_{0}, r_{v}$ and $\alpha$.
We can also get
\begin{equation}
\partial_{r} x=\sqrt{\frac{\frac{e^{2 s r_{0}^{2}}}{r_{0}^{4}} f^{2}\left(r_{0}\right)}{\frac{e^{2 s r^{2}}}{r^{4}} f^{2}(r) f\left(r_{0}\right)-\frac{e^{2 s r_{0}^{2}}}{r_{0}^{4}} f^{2}\left(r_{0}\right) f(r)}},
\end{equation}
and
\begin{equation}\label{22}
\frac{e^{s r_{v}^{2}}}{r_{v}^{2}} f\left(r_{v}\right)-\frac{e^{s r_{0}^{2}}}{r_{0}^{2}} \sqrt{f\left(r_{0}\right)} \sqrt{f\left(r_{v}\right)+\tan ^{2} \alpha}=0.
\end{equation}

At this step, we must find the relationship between $r_{v}$ and $\alpha$.
We can obtain the result through the equilibrium condition at the baryon vertex.
This differs between the three-quark (A) and three-quark (B) models.
The force balance equation at the point $r = r_v$ is

\begin{equation}\label{equillibrium}
\bf{e_{1}+e_{2}+e_{3}+f_{v}}=0.
\end{equation}

\begin{figure}[H]
    \centering
	\includegraphics[width=12cm]{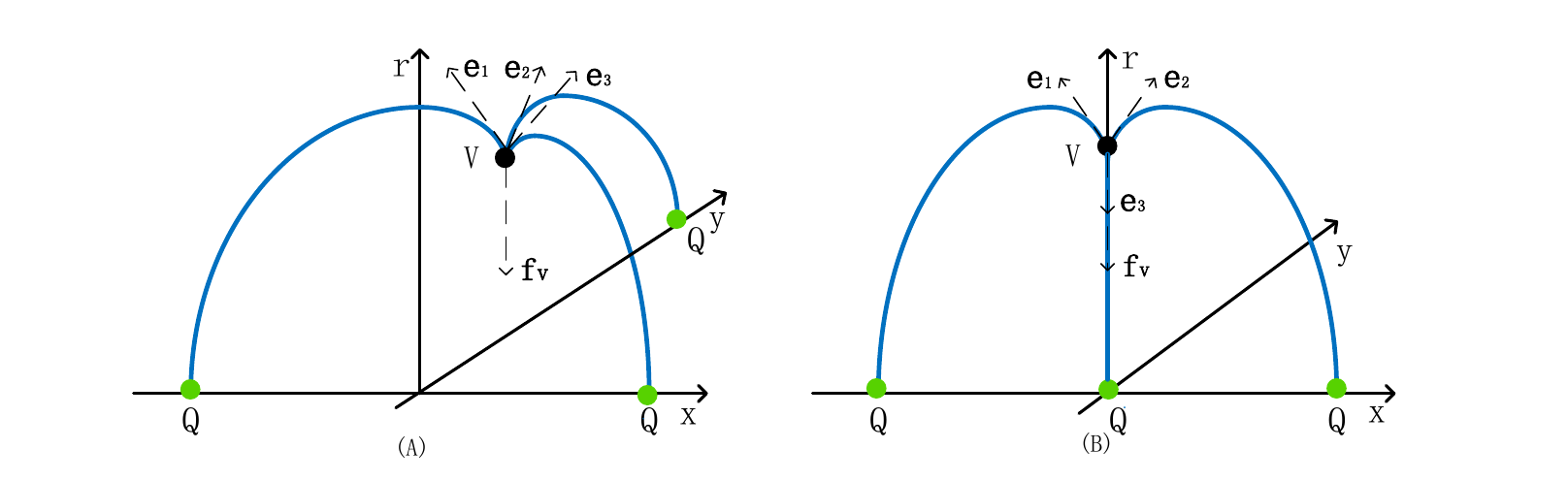}
    \caption{\label{1}(A) The heavy quarks are at the vertices of the equilateral triangle for QQQ. (B) Symmetric collinear geometry for QQQ.}
    \label{fig:1}
\end{figure}

Here, $\bf{e_i}$ is the string tension, and $\bf{f_v}$ is the gravitational force acting on the vertex.
The presence of this force is the main difference between string models in flat spaces and those in curved spaces.
In this model, $\bf{f_v}$ only has one non-zero component in the r-direction.
A formula for this component can be derived from the action $E_{\text {vert }}=\mathrm{T} \cdot S_{\text {vert }}$.
Explicitly, $f_{v}^{r}=-\delta E_{\text {vert }} / \delta r$ and $r$ is the coordinate of the vertex.
The possible configurations of the triply heavy baryon are shown in Fig.\ref{1}.

\subsubsection{QQQ(A) equilateral triangle geometry}

\begin{figure}[H]
    \centering
	\includegraphics[width=6.9cm]{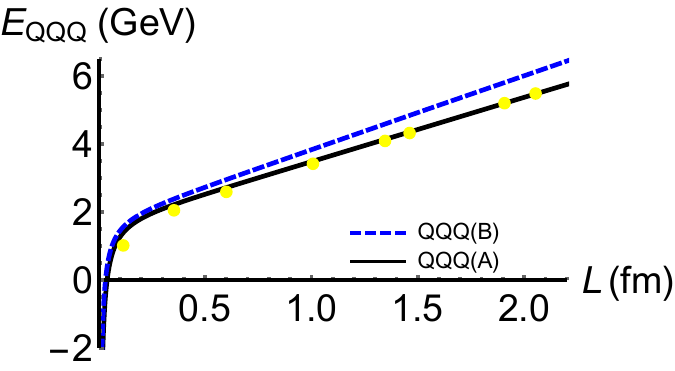}
    \caption{\label{H1dg}The potential energy for three-heavy quarks (heavy baryons) as a function of the separation distance at vanishing temperature. The dots are the results of lattice QCD\cite{Alexandrou:2002sn}. Here $s=0.450$ GeV\cite{Andreev:2006vy}, $\mathrm{g}=0.176$\cite{Bulava:2019iut}, $\mathrm{k}=-0.102$\cite{Greenberg:1981xn} and $\mathrm{c}=0.623$ GeV\cite{Andreev:2021bfg}.}
    \label{fig:H1dg}
\end{figure}

\begin{figure}[H]
    \centering
	\includegraphics[width=12cm]{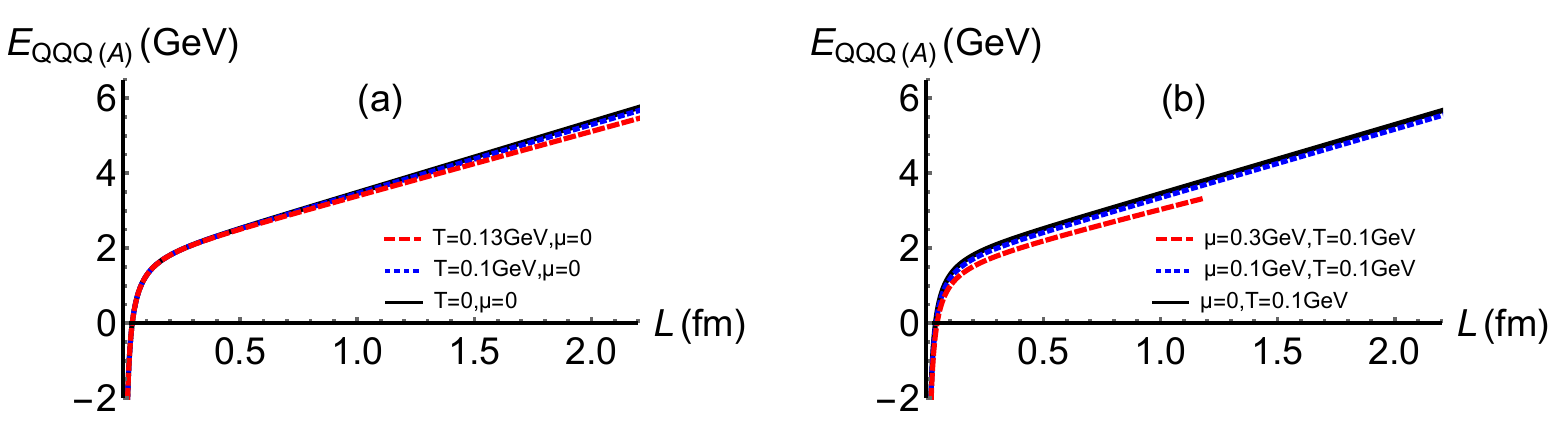}
    \caption{\label{H1dtu}Model A: panel (a) shows the potential energy diagram for different temperatures at a zero chemical potential. Panel (b) presents the potential energy diagram for different chemical potentials at a temperature of $T=0.1$ GeV.}
     \label{fig:H1dtu}
\end{figure}

In the three-quark(A), each force is given by
\begin{equation}
\begin{gathered}
\boldsymbol{\mathrm{f}}_{\mathrm{v}}=\left(0,0,-3 \mathrm{g} \mathrm{k} \partial_{r_{v}} \frac{e^{-2 s r_{v}^{2}}}{r_{v}} \sqrt{f\left(r_{v}\right)}\right) \\
\boldsymbol{e}_{\mathbf{1}}=-\mathrm{g} \frac{e^{s r_{v}^{2}}}{r_{v}^{2}}\Bigg(\cos \beta \frac{f\left(r_{v}\right)}{\sqrt{f\left(r_{v}\right)+\tan ^{2} \alpha}},\sin \beta \frac{f\left(r_{v}\right)}{\sqrt{f\left(r_{v}\right)+\tan ^{2} \alpha}},-\frac{1}{\sqrt{1+f\left(r_{v}\right) \cot ^{2} \alpha}}\Bigg) \\
\boldsymbol{e}_{2}=-\mathrm{g} \frac{e^{s r_{v}^{2}}}{r_{v}^{2}}\Bigg(-\cos \beta \frac{f\left(r_{v}\right)}{\sqrt{f\left(r_{v}\right)+\tan ^{2} \alpha}},\sin \beta \frac{f\left(r_{v}\right)}{\sqrt{f\left(r_{v}\right)+\tan ^{2} \alpha}},-\frac{1}{\sqrt{1+f\left(r_{v}\right) \cot ^{2} \alpha}}\Bigg) \\
\boldsymbol{e}_{3}=-\mathrm{g} \frac{e^{s r_{v}^{2}}}{r_{v}^{2}}\Bigg(0,-\frac{f\left(r_{v}\right)}{\sqrt{f\left(r_{v}\right)+\tan ^{2} \alpha}},-\frac{1}{\sqrt{1+f\left(r_{v}\right) \cot ^{2} \alpha}}\Bigg) \\
\end{gathered}
\end{equation}
Regarding the symmetry in $x$ and $y$, we only need to consider the force balance in the r direction.
The force balance equation leads to
\begin{equation}
\frac{e^{s r_{v}^{2}}}{r_{v}^{2}} \frac{1}{\sqrt{1+f\left(r_{v}\right) \cot ^{2} \alpha}}-\mathrm{k} \partial_{r_{v}} \frac{e^{-2 s r_{v}^{2}}}{r_{v}} \sqrt{f\left(r_{v}\right)}=0.
\end{equation}
Therefore, $L$ and $E$ can be calculated as

\begin{eqnarray}
 L&=&\sqrt{3}\left(\int_{0}^{r_{0}} \partial_{r} x d r+\int_{r_{v}}^{r_{0}} \partial_{r} x d r\right)\\ \notag
&=&\sqrt{3}\int_{0}^{r_{0}} \sqrt{\frac{\frac{e^{2 s r_{0}^{2}}}{r_{0}^{4}} f^{2}\left(r_{0}\right)}{\frac{e^{2 s r^{2}}}{r^{4}} f^{2}(r) f\left(r_{0}\right)-\frac{e^{2 s r_{0}^{2}}}{r_{0}^{4}} f^{2}\left(r_{0}\right) f(r)}} d r +\sqrt{3}\int_{r_{v}}^{r_{0}} \sqrt{\frac{\frac{e^{2 s r_{0}^{2}}}{r_{0}^{4}} f^{2}\left(r_{0}\right)}{\frac{e^{2 s r^{2}}}{r^{4}} f^{2}(r) f\left(r_{0}\right)-\frac{e^{2 s r_{0}^{2}}}{r_{0}^{4}} f^{2}\left(r_{0}\right) f(r)}} d r,
\end{eqnarray}
\begin{equation}
\begin{gathered}
E=3\,\mathrm{g}\left(\int_{0}^{r_{0}} \frac{e^{s r^{2}}}{r^{2}} \sqrt{1+f(r)\left(\partial_{r} x\right)^{2}}-\frac{1}{r^{2}} d r+
\int_{r_{v}}^{r_{0}} \frac{e^{s r^{2}}}{r^{2}} \sqrt{1+f(r)\left(\partial_{r} x\right)^{2}} dr\right)\\
-\frac{3 \mathrm{g}}{r_{0}}+3 \mathrm{g} \mathrm{k} \frac{e^{-2 s r_{v}^{2}}}{r_{v}} \sqrt{f\left(r_{v}\right)}+3 c-\mu.
\end{gathered}
\end{equation}
The potential energy of the three-quark ground state is shown in Fig.~\ref{H1dg}.
And the potential energy of model A at different temperatures and chemical potentials is shown in Fig.~\ref{H1dtu}.

\subsubsection{QQQ(B)symmetric collinear geometry}

\begin{figure}[H]
    \centering
	\includegraphics[width=12cm]{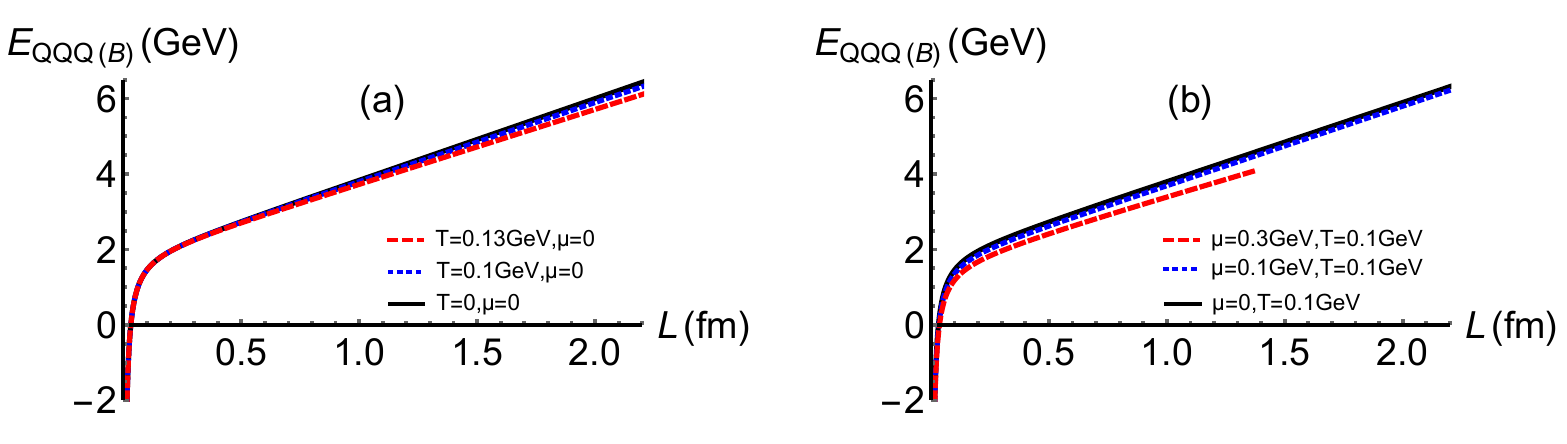}
    \caption{\label{H1gtu}Model B: panel (a) represents the potential energy diagram for different temperatures at zero chemical potential. Panel (b) illustrates the potential energy diagram for different chemical potentials at a temperature of $T=0.1$ GeV.}
    \label{fig:H1gtu}
\end{figure}

In the three-quark(B), each force is given by
\begin{equation}
\begin{gathered}
\boldsymbol{\mathrm{f}}_{\mathrm{v}}=\left(0,-3\, \mathrm{g}\, \mathrm{k} \,\partial_{r_{v}} \frac{e^{-2 s r_{v}^{2}}}{r_{v}} \sqrt{f\left(r_{v}\right)}\right) \\
\boldsymbol{e}_{\mathbf{1}}=-\mathrm{g} \frac{e^{s r_{v}^{2}}}{r_{v}^{2}}\Bigg( \frac{f\left(r_{v}\right)}{\sqrt{f\left(r_{v}\right)+\tan ^{2} \alpha}},-\frac{1}{\sqrt{1+f\left(r_{v}\right) \cot ^{2} \alpha}}\Bigg) \\
\boldsymbol{e}_{2}=-\mathrm{g} \frac{e^{s r_{v}^{2}}}{r_{v}^{2}}\Bigg(- \frac{f\left(r_{v}\right)}{\sqrt{f\left(r_{v}\right)+\tan ^{2} \alpha}},-\frac{1}{\sqrt{1+f\left(r_{v}\right) \cot ^{2} \alpha}}\Bigg) \\
\boldsymbol{e}_{3}=-\mathrm{g} \frac{e^{s r_{v}^{2}}}{r_{v}^{2}}\Bigg(0,1\Bigg) \\
\end{gathered}
\end{equation}
the force balance equation
\begin{equation}
-2 \frac{e^{s r_{v}^{2}}}{r_{v}^{2}} \frac{1}{\sqrt{1+f\left(r_{v}\right) \cot ^{2} \alpha}}+\frac{e^{s r_{v}^{2}}}{r_{v}^{2}}+3\mathrm{k} \partial_{r_{ v}} \frac{e^{-2 s r_{v}^{2}}}{r_{v}} \sqrt{f\left(r_{v}\right)}=0.
\end{equation}
Thus, we can similarly obtain inter-quark distance
\begin{equation}
\begin{gathered}
L=\int_{0}^{r_{0}} \partial_{r} x d r+\int_{r_{v}}^{r_{0}} \partial_{r} x d r \\
=\int_{0}^{r_{0}} \sqrt{\frac{\frac{e^{2 s r_{0}^{2}}}{r_{0}^{4}} f^{2}\left(r_{0}\right)}{\frac{e^{2 s r^{2}}}{r^{4}} f^{2}(r) f\left(r_{0}\right)-\frac{e^{2 s r_{0}^{2}}}{r_{0}^{4}} f^{2}\left(r_{0}\right) f(r)}} d r
+\int_{r_{v}}^{r_{0}} \sqrt{\frac{\frac{e^{2 s r_{0}^{2}}}{r_{0}^{4}} f^{2}\left(r_{0}\right)}{\frac{e^{2 s r^{2}}}{r^{4}} f^{2}(r) f\left(r_{0}\right)-\frac{e^{2 s r_{0}^{2}}}{r_{0}^{4}} f^{2}\left(r_{0}\right) f(r)}} d r,
\end{gathered}
\end{equation}
and the potential energy as in the model A: 
\begin{equation}
\begin{aligned}
\begin{gathered}
E= 2 \mathrm{g}\left(\int_{0}^{r_{0}} \frac{e^{s r^{2}}}{r^{2}} \sqrt{1+f(r)\left(\partial_{r} x\right)^{2}}-\frac{1}{r^{2}} d r+\int_{r_{v}}^{r_{0}} \frac{e^{s r^{2}}}{r^{2}} \sqrt{1+f(r)\left(\partial_{r} x\right)^{2}} d r\right)\\
-\frac{2 \mathrm{g}}{r_{0}}+\mathrm{g} \int_{0}^{r_{v}} \frac{e^{s r^{2}}}{r^{2}}-\frac{1}{r^{2}} d r-\frac{\mathrm{g}}{r_{v}}
+3 \mathrm{g} \mathrm{k} \frac{e^{-2 s r_{v}^{2}}}{r_{v}} \sqrt{f(r_{v})}+3 c-\mu.
\end{gathered}
\end{aligned}
\end{equation}
The potential energy diagram of Model B with different temperatures and chemical potentials is shown in Fig.~\ref{H1gtu}.

\subsection{QQq model}
\begin{figure}[H]
    \centering
	\includegraphics[width=12cm]{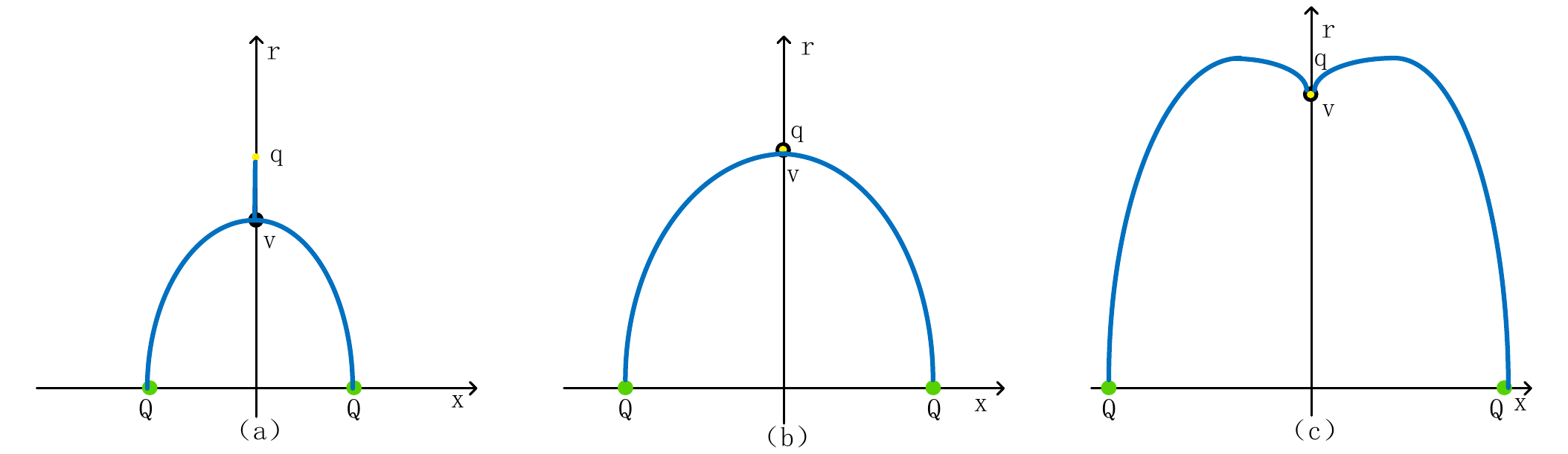}
    \caption{\label{2}According to different string distances, we classify QQq into three distinct states. In the subsequent part, we will discuss the calculation methods for the potential energy of each state. For our model calculations, we adopt the following parameters: $s=0.450$ GeV, $\mathrm{g}=0.176$, $\mathrm{n}=3.057$\cite{Bulava:2019iut}, $\mathrm{k}=-\frac{1}{4}e^{\frac{1}{4}}$\cite{Bulava:2019iut} and $\mathrm{c}=0.623$ GeV. Here, we use different values for the $\mathrm{k}$-value because we are using a specific model\cite{Andreev:2021bfg}.}
    \label{fig:2}
\end{figure}

In this section, we will focus on discussing the cases of QQq.
The configuration of QQq will vary with the distance between the quarks, and for our model, QQq is divided into three cases.
The different configurations of QQq are depicted in Fig.~\ref{2}.

\subsubsection{QQq Small}
In this case, the total effect can be expressed as

\begin{equation}\label{QQq}
S=\sum_{i=1}^3 S_{N G}^{(i)}+S_{\text {vert }}+S_{\mathrm{q}}+2 S_{A}|_{r=0}+S_{A}|_{r=r_{q}}.
\end{equation}
The full action can be written as
\begin{equation}
\begin{aligned}
\begin{gathered}
S=\frac{2 \mathrm{g}}{\mathrm{T}} \int d x \frac{e^{s r^2}}{r^2} \sqrt{f(r)+\left(\partial_x r\right)^2}+\frac{3 \mathrm{k} \mathrm{g}}{\mathrm{T}} \frac{e^{-2 s r_v^2}}{r_v} \sqrt{f(r_{v})}+\frac{\mathrm{n}\mathrm{g} }{\mathrm{T}} \frac{\mathrm{e}^{\frac{1}{2}} \mathrm{~s} r_v^2}{r_v} \sqrt{f\left(r_v\right)}\\
-\frac{2}{3}\frac{A_0(0)}{\mathrm{T}}-\frac{1}{3}\frac{A_0(r_{q})}{\mathrm{T}}.
\end{gathered}
\end{aligned}
\end{equation}
Thus, we have
\begin{equation}
\mathcal{L}=w(r) \sqrt{f(r)+\left(\partial_x r\right)^2},w(r)=\frac{e^{s r^2}}{r^2},
\end{equation}
and
\begin{equation}
\mathcal{L}-r^{\prime} \frac{\partial \mathcal{L}}{\partial r^{\prime}}=\frac{w(r) f(r)}{\sqrt{f(r)+\left(\partial_x r\right)^2}}=\text { constant }
\end{equation}
Using the conservation of energy, we have 
\begin{equation*}
  \frac{w(r) f(r)}{\sqrt{f(r)+\left(\partial_x r\right)^2}}=\frac{w\left(r_v\right)f\left(r_v\right)}{\sqrt{f\left(r_v\right)+\tan ^2 \alpha}}  
\end{equation*}
at the $U$-shape string maximum point $r_0$. Thus, we get
\begin{equation}
\partial_r x=\sqrt{\frac{\omega\left(r_v\right)^2 f\left(r_v\right)^2}{\left(f\left(r_\nu\right)+\tan ^2 \alpha\right) \omega(r)^2 f(r)^2-f(r) w\left(r_v\right)^2 f\left(r_v\right)^2}}.
\end{equation}

For the force equilibrium in the model,

\begin{equation*}
    \mathbf{f}_\mathrm{v}=\left(0,-3 \mathbf{\mathrm{g}} \mathrm{k} \partial_{r_v}\left(\frac{\mathrm{e}^{-2 s r_v^2}}{r_v} \sqrt{f\left(r_v\right)}\right)\right),
\end{equation*}

and string tensions are 

\begin{eqnarray*}
\mathbf{e}_1&=&\mathbf{\mathrm{g}} w\left(r_v\right)\left(-\frac{f\left(r_v\right)}{\sqrt{\tan ^2 \alpha+f\left(r_v\right)}},-\frac{1}{\sqrt{f\left(r_v\right) \cot ^2 \alpha+1}}\right),\\
\mathbf{e}_2&=&\mathbf{\mathrm{g}} w\left(r_v\right)\left(\frac{f\left(r_v\right)}{\sqrt{\tan ^2 \alpha+f\left(r_v\right)}},-\frac{1}{\sqrt{f\left(r_v\right) \cot ^2 \alpha+1}}\right),\\
\mathbf{e}_3&=&\mathbf{\mathrm{g}} w\left(r_v\right)(0,1).
\end{eqnarray*}


Therefore, after imposing force equilibrium at the vertex

\begin{equation}
\mathbf{e}_1+\mathbf{e}_2+\mathbf{e}_3+\mathbf{f}_\mathrm{v}=0
\end{equation}

we obtain the force balance equation given by

\begin{equation}
2 \mathrm{g} \frac{e^{s r_v^2}}{r_v^2} \frac{1}{\sqrt{1+f\left(r_v\right)\cot^2 \alpha}}-\mathrm{g} \frac{e^{s r_v^2}}{r_v^2}+3 \mathrm{g} \mathrm{k}\left(\frac{e^{-2 s r_{v}^2}}{r_v} \sqrt{f\left(r_v\right)}\right)^{\prime}=0
\end{equation}

As the string distance increases, the baryon vertex $\rm v$ also arises. The limiting case for the first state is when the baryon vertex $v$ and the light quark $\rm q$ are heavy. The limiting case of the first configuration is that the position of the light quark reaches the vertex one. We can figure it out by the following formula.

\begin{equation}
\mathrm{g} \frac{e^{s r_q^2}}{r_q}+\mathrm{n} \mathrm{g}\left(\frac{e^{\frac{1}{2} s r_q^2}}{r_q} \sqrt{f\left(r_q\right)}\right)^{\prime} -\left(\frac{1}{3}A_0\left(r_q\right)\right)^{\prime}=0
\end{equation}

The first term is the string tension, the second is the force of the light quark, and the third is the force of the world-sheet action.

Therefore, $L$ and $E$ can be calculated as
\begin{equation}
\begin{aligned}
\begin{gathered}
L=2 \int_0^{r_v} \partial_r x d r \\
=2 \int_0^{r_v} \sqrt{\frac{w^2\left(r_v\right) f^2\left(r_v\right)}{w^2(r) f^2(r)\left(f\left(r_v\right)+\tan ^2 \alpha\right)-w^2\left(r_v\right) f^2\left(r_v\right) f(r)}} d r ,
\end{gathered}
\end{aligned}
\end{equation}
and
\begin{equation}
\begin{aligned}
\begin{gathered}
E=2 \mathrm{g} \int_0^{r_v} w(r)\sqrt{1+f\left(r\right)\frac{w^2\left(r_v\right) f^2\left(r_v\right)}{w^2(r) f^2(r)\left(f\left(r_v\right)+\tan ^2 \alpha\right)-w^2\left(r_v\right) f^2\left(r_v\right) f(r)}}-\frac{1}{r^2} d r\\
-2\frac{\mathrm{g}}{r_{v}}+\mathrm{g} \int_{r_v}^{r_q} w(r)d r
+3 \mathrm{g} \mathrm{k} \frac{e^{-2 s r_v^2}}{r_v} \sqrt{f\left(r_v\right)}+\mathrm{n} \mathrm{g} \frac{\mathrm{e}^{\frac{1}{2} s r_q^2}}{r_q} \sqrt{f\left(r_q\right)}-\frac{2}{3}A_{0}(0)-\frac{1}{3}A_{0}({r_{q}})+2c.
\end{gathered}
\end{aligned}
\end{equation}

\begin{figure}[H]
    \centering
	\includegraphics[width=17cm]{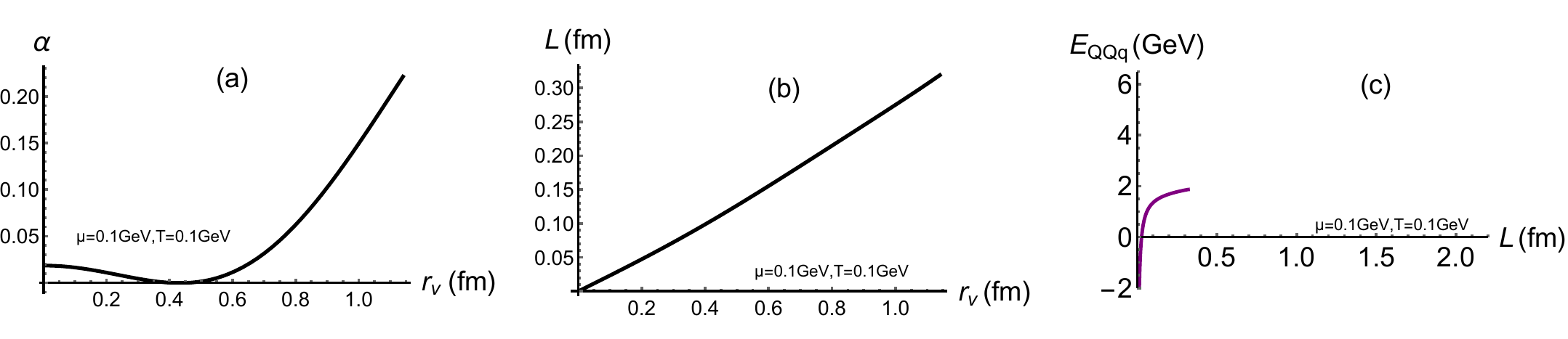}
    \caption{\label{H2u01t01s} (a) $\alpha$ as a function of $r_{v}$. (b) Separation distance $L$ as a function of $r_{v}$. (c) The energy $E$ as a function of $L$ at $\mu=0.1$ GeV,$T=0.1$ GeV.}
    \label{fig:H2u01t01s}
\end{figure}

 The plots for $\alpha$, $L$, and $E$ for the QQq Small configuration are shown in  Fig.~\ref{H2u01t01s}.
 
\subsubsection{QQq Intermediate}
The total action in this configuration reads as

\begin{equation}\label{QQq-int}
S=\sum_{i=1}^2 S_{N G}^{(i)}+S_{\text {vert }}+S_{\mathrm{q}}+2 S_{A}|_{r=0}+S_{A}|_{r=r_{q}},
\end{equation}

that can be explicitly written as 

\begin{equation}
\begin{aligned}
\begin{gathered}
S=\frac{2 \mathrm{g}}{\mathrm{T}} \int d x \frac{e^{s r^2}}{r^2} \sqrt{f(r)+\left(\partial_x r\right)^2}+\frac{3 \mathrm{k} \mathrm{g}}{\mathrm{T}} \frac{e^{-2 s r_v^2}}{r_v} \sqrt{f(r_{v})}+\frac{\mathrm{n}\mathrm{g} }{\mathrm{T}} \frac{\mathrm{e}^{\frac{1}{2}} \mathrm{~s} r_v^2}{r_v} \sqrt{f\left(r_v\right)}\\
-\frac{2}{3}\frac{A_0(0)}{\mathrm{T}}-\frac{1}{3}\frac{A_0(r_{q})}{\mathrm{T}}.
\end{gathered}
\end{aligned}
\end{equation}

The force balance equation at the point $r=r_v$ is
\begin{equation}
\mathbf{e}_1+\mathbf{e}_2+\mathbf{f}_\mathrm{v}+\mathbf{f}_\mathrm{q}+\mathbf{f}_A=0,
\end{equation}
where each force is given by
\begin{equation}
\begin{aligned}
\begin{gathered}
\begin{aligned}
&\mathbf{f}_A=\left(0, \partial_{r_q}\left(A_0\left(r_q\right)\right)\right. \\
&\mathbf{f}_\mathrm{q}=\left(0,-\mathrm{n} \mathrm{g} \partial_{r_q}\left(\frac{e^{\frac{1}{2} s r_q^2}}{r_q} \sqrt{f\left(r_q\right)}\right)\right. \\
&\mathbf{f}_\mathrm{v}=\left(0,-3 \mathrm{g} \mathrm{k} \partial_{r_v}\left(\frac{e^{-2s r_v^2}}{r_v} \sqrt{f\left(r_v\right)}\right)\right. \\
&\mathbf{e}_1=\mathrm{g} w\left(r_v\right)\left(-\frac{f\left(r_v\right)}{\sqrt{\tan ^2 \alpha+f\left(r_v\right)}},-\frac{1}{\sqrt{f\left(r_v\right)\left(cot^2 \alpha+1\right)}}\right) \\
&\mathbf{e}_2=\mathrm{g} w\left(r_v\right)\left(\frac{f\left(r_v\right)}{\sqrt{\tan ^2 \alpha+f\left(r_v\right)}},-\frac{1}{\sqrt{f\left(r_v\right)\left(cot ^2 \alpha+1\right)}}\right)
\end{aligned}
\end{gathered}
\end{aligned}
\end{equation}
with $r_{\mathrm{q}} = r_{\mathrm{v}}$. The force balance equation leads to
\begin{equation}
\begin{aligned}
\begin{gathered}
2 \mathrm{g} \frac{e^{s r_v^2}}{r_v^2} \frac{1}{\sqrt{\left(1+f\left(r_v\right)\right) cot^2 \alpha}}+3 \mathrm{g} \mathrm{k} \partial_{r_v}\left(\frac{e^{-2 s r_v^2}}{r_v} \sqrt{f\left(r_v\right)}\right)\\
+\mathrm{n} \mathrm{g} \partial_{r_v}\left(\frac{e^{\frac{1}{2} s r_v^2}}{r_v} \sqrt{f\left(r_v\right)}\right)-\partial_{r_v}\left(A_v\left(r_v\right)\right)=0
\end{gathered}
\end{aligned}
\end{equation}
The QQq becomes the second configuration, where the baryon vertices coincide with the light quarks. As the string distance increases, the baryon vertices and the light quarks move away from the heavy quarks. The angle $\alpha$ will tend to zero in the second configuration. When $\alpha$ tends to be negative, the third state has to start. Now, the force equilibrium becomes
\begin{equation}
3 \mathrm{g} \mathrm{k} \partial_{r_v}\left(\frac{e^{-2 s r_v^2}}{r_v} \sqrt{f\left(r_v\right)}\right)+\mathrm{n} \mathrm{g} \partial_{r_v}\left(\frac{e^{\frac{1}{2} s r_v^2}}{r_v} \sqrt{f\left(r_v\right)}\right)-\partial_{r_v}\left(A_v\left(r_v\right)\right)=0
\end{equation}
The solution to the equation is the limiting position of the second state.

Therefore, $L$ and $E$ can be calculated as
\begin{equation}
\begin{aligned}
\begin{gathered}
L=2 \int_0^{r_v} \partial_r x d r \\
=2 \int_0^{r_v} \sqrt{\frac{w^2\left(r_v\right) f^2\left(r_v\right)}{w^2(r)\left(f\left(r_v\right)+\tan ^2 \alpha\right) f^2(r)-f(r) w^2\left(r_v\right) f^2\left(r_v\right)}} d r
\end{gathered}
\end{aligned}
\end{equation}
and
\begin{equation}
\begin{aligned}
\begin{gathered}
E=2 \mathrm{g} \int_0^{r_v} w(r) \sqrt{1+f\left(r\right)\left(\partial_r x\right)^2}-\frac{1}{r^2} d r-2\frac{\mathrm{g}}{r_v}+\mathrm{n} \mathrm{g} \frac{e^{\frac{1}{2} s r_v^2}}{r_v} \sqrt{f\left(r_v\right)}\\
+3 \mathrm{g} \mathrm{k} \frac{e^{-2 s r_v^2}}{r_v} \sqrt{f\left(r_v\right)}-\frac{2}{3} A_0(0)-\frac{1}{3} A_0\left(r_q\right)+2 c
\end{gathered}
\end{aligned}
\end{equation}

\begin{figure}[H]
    \centering
	\includegraphics[width=17cm]{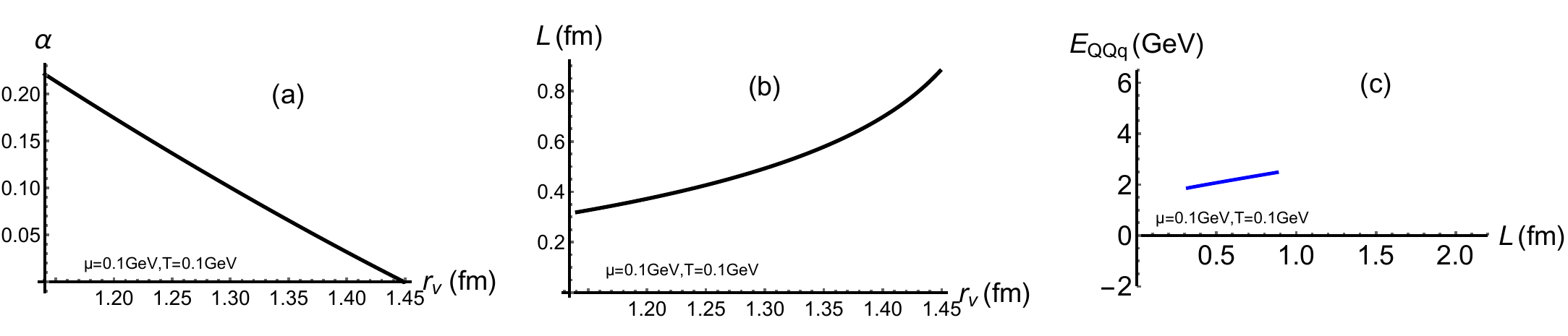}
    \caption{\label{H2u01t01i}(a)$\alpha$ as a function of $r_{\mathrm{v}}$. (b)separate distance $L$ as a function of $r_{\mathrm{v}}$. (c)the energy $E$ as a function of $L$.at $\mu$=0.1GeV,$T$=0.1GeV.}
    \label{fig:H2u01t01i}
\end{figure}

Plots for $\alpha$, $L$, and $E$ of QQq intermediate configuration are shown in  Fig.~\ref{H2u01t01i}.

\subsubsection{QQq Large}
The total action for this configuration is the same as Eq. \eqref{QQq-int}. Thus, the total action becomes
\begin{equation}
\begin{aligned}
\begin{gathered}
S=\frac{2 \mathrm{g}}{\mathrm{T}} \int d x \frac{e^{s r^2}}{r^2} \sqrt{f(r)+(\partial_{x} r)^2}+\frac{\mathrm{n} \mathrm{g}}{\mathrm{T}} \frac{e^{\frac{1}{2} s r_v^2}}{r_v} \sqrt{f\left(r_v\right)}+\frac{3 \mathrm{g}}{\mathrm{T}} \mathrm{k} \frac{e^{-2 s r_v^2}}{r_v} \sqrt{f\left(r_v\right)}\\
-\frac{2}{3} \frac{ A_{0}(0)}{\mathrm{T}}-\frac{1}{3} \frac{A_0\left(r_q\right)}{\mathrm{T}}
\end{gathered}
\end{aligned}
\end{equation}
The Lagrangian density is written as
\begin{equation}
\mathcal{L}=\frac{e^{s r^{2}}}{r^{2}} \sqrt{f(r)+\left(\partial_{x} r\right)^{2}},
\end{equation}
with the constrain
\begin{equation}
\mathcal{L}-r^{\prime} \frac{\partial\mathcal{L}}{\partial r^{\prime}}=\frac{\frac{e^{s r^{2}}}{r^{2}} f(r)}{\sqrt{f(r)+\left(\partial_{x} r\right)^{2}}}=\rm constant.
\end{equation}
At the points $r_0$ and $r_v$, we have
\begin{equation}\label{20}
\begin{aligned}
&\frac{\frac{e^{s r^{2}}}{r^{2}} f(r)}{\sqrt{f(r)+\left(\partial_{x} r\right)^{2}}}=\frac{e^{s r_{0}^{2}}}{r_{0}^{2}} \sqrt{f\left(r_{0}\right)} \\
&\frac{\frac{e^{s r_{v}^{2}}}{r_{v}^{2}} f\left(r_{v}\right)}{\sqrt{f\left(r_{v}\right)+\tan ^{2} \alpha}}=\frac{e^{s r_{0}^{2}}}{r_{0}^{2}} \sqrt{f\left(r_{0}\right)}.
\end{aligned}
\end{equation}

The above equations can determine the relation of $r_{0}, r_{v}$ and $\alpha$. We can also obtain

\begin{equation}
\partial_{r} x=\sqrt{\frac{\frac{e^{2 s r_{0}^{2}}}{r_{0}^{4}} f^{2}\left(r_{0}\right)}{\frac{e^{2 s r^{2}}}{r^{4}} f^{2}(r) f\left(r_{0}\right)-\frac{e^{2 s r_{0}^{2}}}{r_{0}^{4}} f^{2}\left(r_{0}\right) f(r)}}
\end{equation}
and
\begin{equation}\label{22}
\frac{e^{s r_{v}^{2}}}{r_{v}^{2}} f\left(r_{v}\right)-\frac{e^{s r_{0}^{2}}}{r_{0}^{2}} \sqrt{f\left(r_{0}\right)} \sqrt{f\left(r_{v}\right)+\tan ^{2} \alpha}=0.
\end{equation}

We can obtain the relationship between the highest point $r_0$, the baryon vertex, and the position $r_v$ of the light quark. The result is depicted in Fig.~\ref{H2u01t01r0}.

The force balance equation at the point $r_\mathrm{q} = r_v$ is
\begin{equation}
\mathbf{e}_1+\mathbf{e}_2+\mathbf{f}_\mathrm{v}+\mathbf{f}_\mathrm{q}+\mathbf{f}_A=0.
\end{equation}
Each force is given by
\begin{equation}
\begin{aligned}
\begin{gathered}
\begin{aligned}
&\mathbf{f}_A=\left(0, \partial_{r_q}\left(A_0\left(r_q\right)\right)\right. \\
&\mathbf{f}_\mathrm{q}=\left(0,-\mathrm{n} \mathrm{g} \partial_{r_q}\left(\frac{e^{\frac{1}{2} s r_q^2}}{r_q} \sqrt{f\left(r_q\right)}\right)\right. \\
&\mathbf{f}_\mathrm{v}=\left(0,-3 \mathrm{g} \mathrm{k} \partial_{r_v}\left(\frac{e^{-2s r_v^2}}{r_v} \sqrt{f\left(r_v\right)}\right)\right. \\
&\mathbf{e}_1=\mathrm{g} w\left(r_v\right)\left(-\frac{f\left(r_v\right)}{\sqrt{\tan ^2 \alpha+f\left(r_v\right)}},\frac{1}{\sqrt{f\left(r_v\right)\left(cot^2 \alpha+1\right)}}\right) \\
&\mathbf{e}_2=\mathrm{g} w\left(r_v\right)\left(\frac{f\left(r_v\right)}{\sqrt{\tan ^2 \alpha+f\left(r_v\right)}},\frac{1}{\sqrt{f\left(r_v\right)\left(cot ^2 \alpha+1\right)}}\right)
\end{aligned}
\end{gathered}
\end{aligned}
\end{equation}
The force balance equation leads to
\begin{equation}
\begin{aligned}
\begin{gathered}
-2 \mathrm{g} \frac{e^{s r_v^2}}{r_v^2} \frac{1}{\sqrt{\left(1+f\left(r_v\right)\right) cot^2 \alpha}}+3 \mathrm{g} \mathrm{k} \partial_{r_v}\left(\frac{e^{-2 s r_v^2}}{r_v} \sqrt{f\left(r_v\right)}\right)\\
+\mathrm{n} \mathrm{g} \partial_{r_v}\left(\frac{e^{\frac{1}{2} s r_v^2}}{r_v} \sqrt{f\left(r_v\right)}\right)-\partial_{r_v}\left(A_v\left(r_v\right)\right)=0
\end{gathered}
\end{aligned}
\end{equation}
\begin{figure}[H]
    \centering
	\includegraphics[width=6cm]{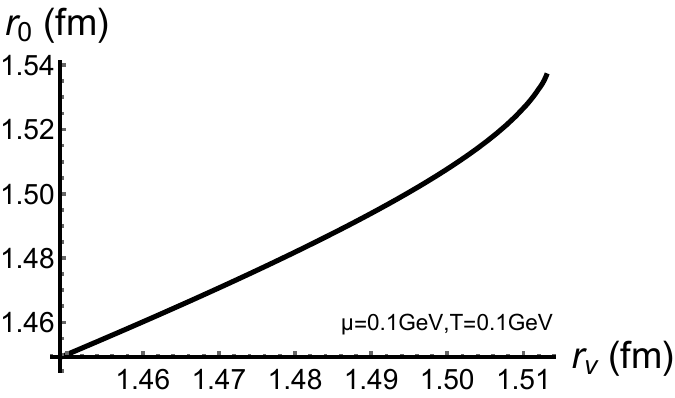}
    \caption{\label{H2u01t01r0}$r_0$ as a function of $r_v$ at $\mu=0.1$ GeV and $T=0.1$ GeV.}
    \label{H2u01t01r0}
\end{figure}
Therefore, $L$ and $E$ can be calculated as
\begin{equation}
\begin{aligned}
\begin{gathered}
L=2 \left(\int_{0}^{r_{0}} \partial_{r} x d r+\int_{r_{v}}^{r_{0}} \partial_{r} x d r\right)\\
=2 \int_{0}^{r_{0}} \sqrt{\frac{\frac{e^{2 s r_{0}^{2}}}{r_{0}^{4}} f^{2}\left(r_{0}\right)}{\frac{e^{2 s r^{2}}}{r^{4}} f^{2}(r) f\left(r_{0}\right)-\frac{e^{2 s r_{0}^{2}}}{r_{0}^{4}} f^{2}\left(r_{0}\right) f(r)}} d r\\
+2 \int_{r_{v}}^{r_{0}} \sqrt{\frac{\frac{e^{2 s r_{0}^{2}}}{r_{0}^{4}} f^{2}\left(r_{0}\right)}{\frac{e^{2 s r^{2}}}{r^{4}} f^{2}(r) f\left(r_{0}\right)-\frac{e^{2 s r_{0}^{2}}}{r_{0}^{4}} f^{2}\left(r_{0}\right) f(r)}} d r,
\end{gathered}
\end{aligned}
\end{equation}
\begin{equation}
\begin{gathered}
E=2 \mathrm{g}\left(\int _ { 0 } ^ { r _ { 0 } } w\left(r\right) \sqrt{1+f\left(r\right) \partial r^2 x}-\frac{1}{r^2} d r-\frac{1}{r_0}+\int_{r_v}^{r_0} w\left(r\right) \sqrt{1+f\left(r\right) \partial r^2 x} d r\right)\\
+\mathrm{n} \mathrm{g} \frac{e^{\frac{1}{2} s r_v^2}}{r_v} \sqrt{f\left(r_v\right)}
+3 \mathrm{g} \mathrm{k} \frac{e^{-2 s r_v^2}}{r_v} \sqrt{f\left(r_v\right)}-\frac{1}{3} A_0\left(r_q\right)-\frac{2}{3} A_0(0)+2 c .
\end{gathered}
\end{equation}

\begin{figure}[H]
    \centering
	\includegraphics[width=17cm]{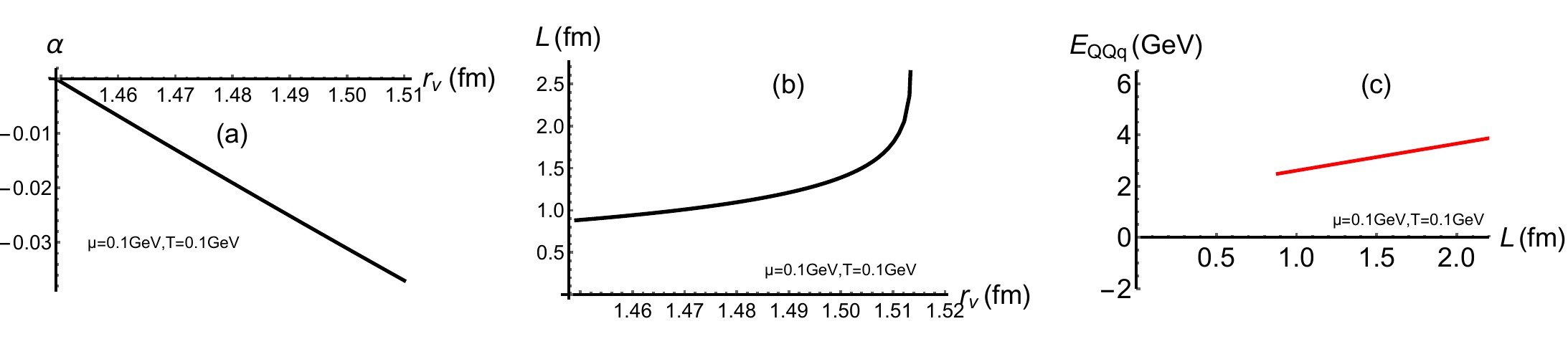}
    \caption{\label{H2u01t01l}Panel (a): $\alpha$ as a function of $r_{v}$. Panel (b): separation distance $L$ as a function of $r_{v}$. Panel (c): the energy $E$ as a function of $L$ at $\mu=0.1$ GeV, $T=0.1$ GeV.}
    \label{fig:H2u01t01l}
\end{figure}

\begin{figure}[H]
    \centering
	\includegraphics[width=15cm]{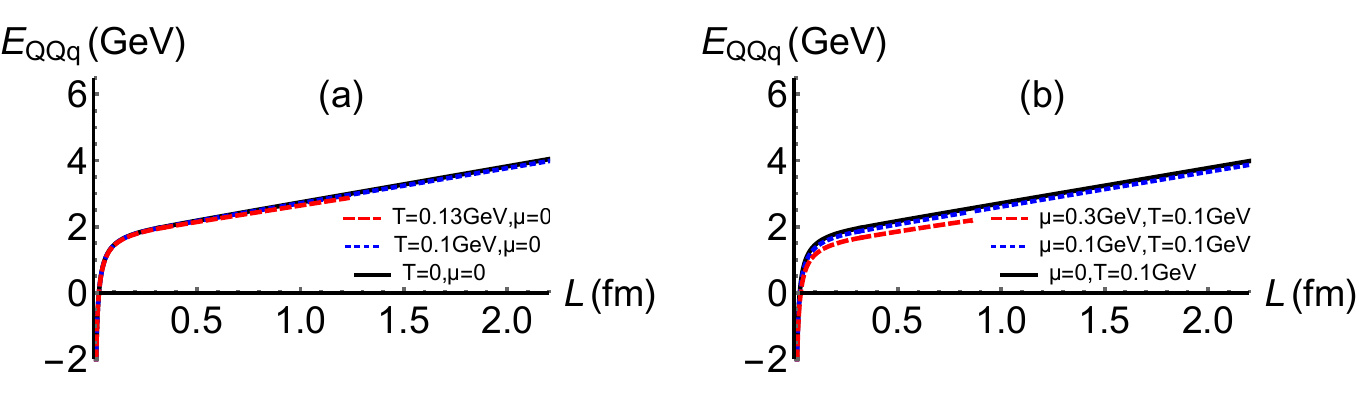}
    \caption{\label{H2tu}QQq model: (a) The potential energy diagram for different temperatures at a chemical potential $\mu=0$. (b) The potential energy diagram for different chemical potentials at a temperature of $T=0.1$ GeV.}
    \label{fig:H2tu}
\end{figure}

Plot for $\alpha$, $L$, and $E$ for the QQq large are shown in  Fig.~\ref{H2u01t01l}.
Finally, we get the potential energy of QQq in Fig.~\ref{H2tu}.
\subsection{Q$\bar{q}$, Qqq, qqq models}
\begin{figure}[H]
    \centering
	\includegraphics[width=12cm]{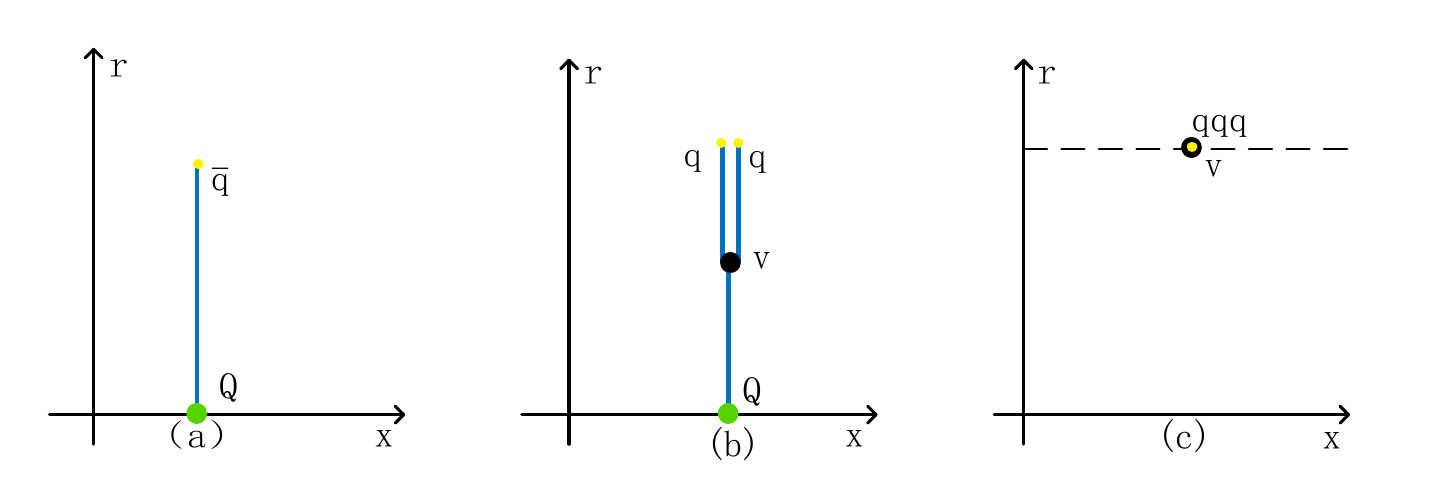}
    \caption{\label{3} Panel (a) is the configuration of Q$\rm \bar q$. Panel (b) is the configuration of Qqq. Panel (c) is the configuration of qqq. Here $\mathrm{s}=0.450$ GeV,$\mathrm{g}=0.176$,$\mathrm{k}=-\frac{1}{4}e^{\frac{1}{4}}$, and $\mathrm{c}=0.623$ GeV.}
    \label{fig:3}
\end{figure}
Now, let us examine the potential energy of the other three models. To do this, we must determine the positions of the light quarks q and $\bar{\mathrm{q}}$. We can achieve this by utilizing the following three force equilibrium equations. The different configurations are shown in Fig.~\ref{3}. For deeper explanations,  see\cite{Andreev:2016hxm, Andreev:2009zk, Andreev:2006eh} whose conventions we generally follow.

In Q$\rm \bar q$ models, we get the force balance equation is
\begin{equation}
\mathrm{g} \frac{e^{s r_{\bar{q}}}}{r_{\bar{q}}}+\mathrm{n} \mathrm{g}\left(\frac{e^{\frac{1}{2} s r_{\bar{q}}^2}}{r_{\bar{q}}} \sqrt{f\left(r_{\bar{q}}\right)}\right)^{\prime}+\left(\frac{1}{3}A_0\left(r_{\bar{q}}\right)\right)^{\prime}=0
\end{equation}
and the we can get the $r_{\bar{q}}$.

Then, in the Qqq models, we get the force balance equation $r_{v}$ and $r_{q}$
\begin{equation}
\mathrm{g} \frac{e^{s r_v^2}}{r_v^2}-3 \mathrm{g} \mathrm{k}\left(\frac{e^{-2s r_v^2}}{r_v} \sqrt{f\left(r_v\right)}\right)^{\prime}=0
\end{equation}
and
\begin{equation}
\mathrm{g} \frac{e^{s r_q^2}}{r_q}+\mathrm{n} \mathrm{g}\left(\frac{e^{\frac{1}{2} s r_q^2}}{r_q} \sqrt{f\left(r_q\right)}\right)^{\prime}-\left(\frac{1}{3}A_0\left(r_q\right)\right)^{\prime}=0
\end{equation}

Finally, in qqq models,$r_{q}$ force balance is
\begin{equation}
3 \mathrm{g} \mathrm{k}\left(\frac{e^{-2 s r_q^2}}{r_q} \sqrt{f\left(r_q\right)}\right)^{\prime}+3 \mathrm{n} \mathrm{g}\left(\frac{e^{\frac{1}{2} s r_q^2}}{r_q} \sqrt{f\left(r_q\right)}\right)^{\prime}-\left(A_0\left(r_q\right)\right)^{\prime}=0
\end{equation}
The potential energy of Q$\bar q$, Qqq, and qqq are
\begin{equation}
E_{Q \bar{q}}=\mathrm{g} \int_0^{r_{\bar{q}}} \frac{e^{s r^2}}{r^2}-\frac{1}{r^2} d r-\frac{\mathrm{g}}{r_{\bar{q}}}+\mathrm{n} \mathrm{g} \frac{e^{\frac{1}{2} s r_{\bar{q}}^2}}{r_{\bar{q}}}+\frac{1}{3} A_0(0)-\frac{1}{3} A_0\left(r_{\bar{q}}\right)+c
\end{equation}
\begin{equation}
\begin{aligned}
\begin{gathered}
E_{Q q q}= \mathrm{g} \int_0^{r_v} \frac{e^{s r^2}}{r^2}-\frac{1}{r^2} d r-\frac{\mathrm{g}}{r_v}+2 \mathrm{g} \int_{r_v}^{r_q} \frac{e^{s r^2}}{r^2} d r+2 \mathrm{n} \mathrm{g} \frac{e^{\frac{1}{2} s r_q{ }^2}}{r_q}+3 \mathrm{g} \mathrm{k} \frac{e^{-2 s r_v^2}}{r_v}\\
-\frac{2}{3} A_0\left(r_q\right)-\frac{1}{3} A_0(0)+ c
\end{gathered}
\end{aligned}
\end{equation}
\begin{equation}
\begin{gathered}
E_{q q q}=
3 \mathrm{g} \mathrm{k} \frac{e^{-2 s r_q^2}}{r_q}+3 \mathrm{n} \mathrm{g} \frac{e^{\frac{1}{2} s r_q^2}}{r_q}-A_0\left(r_q\right)
\end{gathered}
\end{equation}
Therefore, we can compute the potential energy in case 3 and case 4. These energies can be calculated as
\begin{equation}
\begin{gathered}
E_{Q q q}+2 E_{Q \bar{q}}= \\
\mathrm{g} \int_0^{r_v} \frac{e^{s r^2}}{r^2}-\frac{1}{r^2} d r-\frac{\mathrm{g}}{r_v}+2 \mathrm{g} \int_{r_r}^{r_q} \frac{e^{s r^2}}{r^2} d r+2 \mathrm{n} \mathrm{g} \frac{e^{\frac{1}{2} s r_q{ }^2}}{r_q}+3 \mathrm{g} \mathrm{k} \frac{e^{-2 s r_v^2}}{r_v}-\frac{2}{3} A_0\left(r_q\right)-\frac{1}{3} A_0(0) \\
+2 \mathrm{g} \int_0^{r_{\bar{q}}} \frac{e^{s r^2}}{r^2}-\frac{1}{r^2} d r-2 \frac{\mathrm{g}}{r_{\bar{q}}}+2 \mathrm{n} \mathrm{g} \frac{e^{\frac{1}{2} s r_{\bar q}^2}}{r_{\bar{q}}}+\frac{2}{3} A_0\left(r_{\bar{q}}\right)-\frac{2}{3} A_0(0)+3 c
\end{gathered}
\end{equation}
\begin{equation}
\begin{gathered}
E_{q q q}+3 E_{Q_q}= \\
3 \mathrm{g} \mathrm{k} \frac{e^{-2 s r_q^2}}{r_q}+3 \mathrm{n} \mathrm{g} \frac{e^{\frac{1}{2} s r_q^2}}{r_q}-A_0\left(r_q\right) \\
+3 \mathrm{g}\int_0 ^{r_{\bar{q}}} \frac{e^{s r^2}}{r^2}-\frac{1}{r^2} d r-3 \frac{\mathrm{g}}{r_{\bar{q}}}+3 \mathrm{n} \mathrm{g} \frac{e^{\frac{1}{2} s r_{\bar{q}}^2}}{r_{\bar{q}}}+A_0(0)-A_0\left(r_{\bar{q}}\right)+3c
\end{gathered}
\end{equation}

\section{Numerical results }\label{Numerical results}
\begin{figure}[H]
    \centering
	\includegraphics[width=15cm]{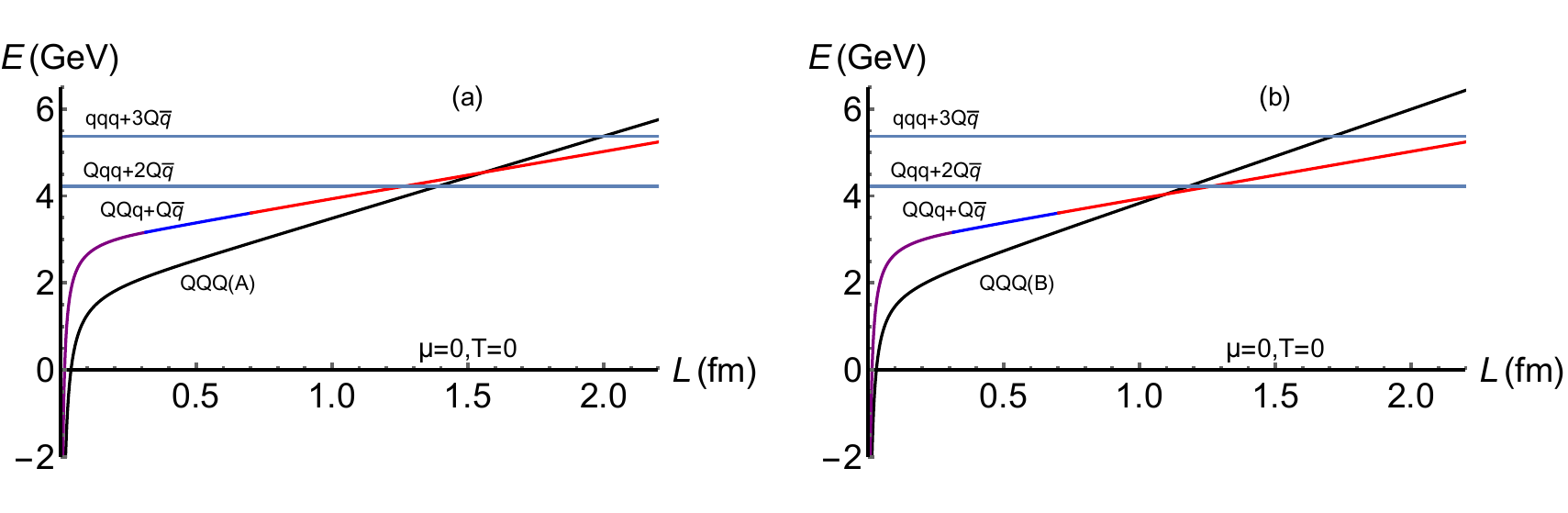}
    \caption{\label{t0u0}Under the conditions $T=0$ and $\mu=0$, we plot the potential energy graphs of the quark states in four cases as a function of the distance between quarks. Panel (a) corresponds to the equilateral triangle geometry, and (b) corresponds to the symmetric collinear geometry.}
    \label{fig:t0u0}
\end{figure}

We calculated the potential energy diagram for the ground state of triply heavy baryons, as shown in Fig.~\ref{t0u0}. This ground state potential energy diagram is obtained under the conditions $T=0$ and $\mu=0$.

\begin{figure}[H]
    \centering
	\includegraphics[width=15cm]{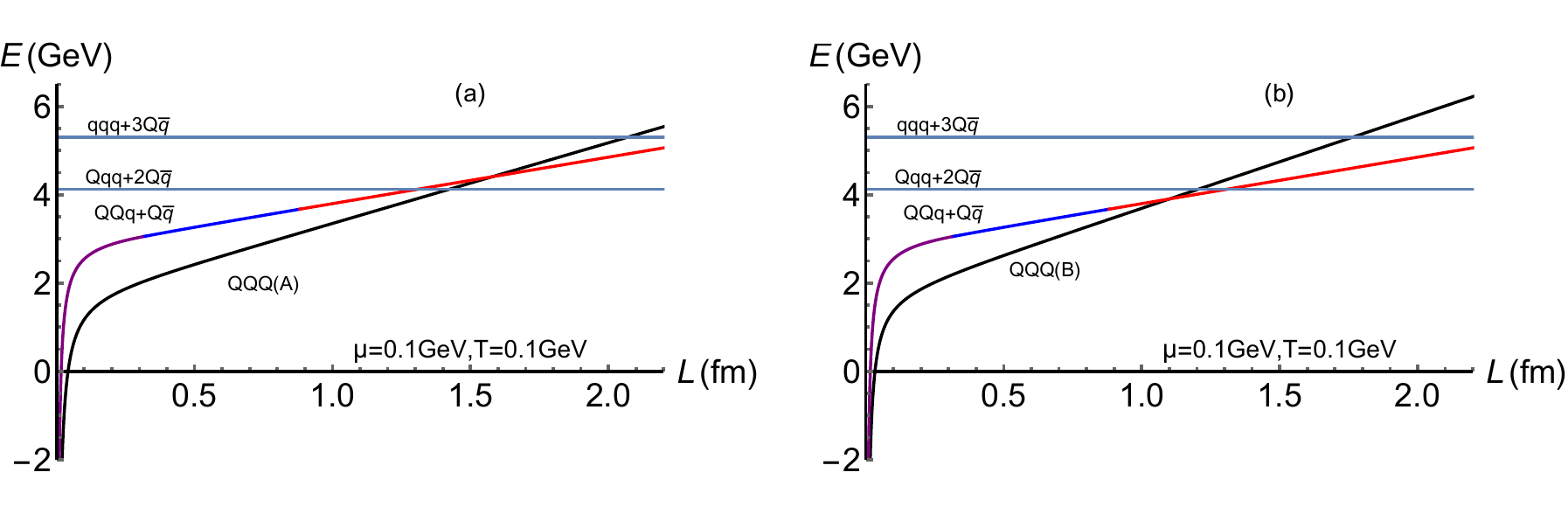}
    \caption{\label{t01u01}By varying the temperature and chemical potential, we can analyze the impact of each one separately to obtain the fragmentation distance. In this context, panel (a) corresponds to the equilateral triangle geometry, and (b) corresponds to the symmetric collinear geometry.}
    \label{fig:t01u01}
\end{figure}

After analyzing the three-quark ground state, we calculated the potential energy diagrams for the finite temperature and finite chemical potential case. These diagrams are depicted in Fig.~\ref{t01u01}.

With these results, we can focus our study on two geometric configurations of the triply heavy quark system: QQQ(A) with an equilateral triangular geometry and QQQ(B) with a symmetric collinear geometry. The following contents will discuss the distinct patterns of string-breaking distances exhibited under these two configurations separately as the thermodynamic conditions are modulated.

Based on the potential energy functions of inter-quark distances presented in Fig.~\ref{t0u0} and Fig.~\ref{t01u01}, we have concluded that for the QQQ(A) configuration with equilateral triangular geometry, the most likely decay mode is into Qqq and 2Q$\rm\bar{q}$, while for the QQQ(B) configuration with symmetric collinear geometry, the most likely decay mode is into QQq and Q$\rm\bar{q}$.
\begin{figure}[H]
    \centering
	\includegraphics[width=12cm]{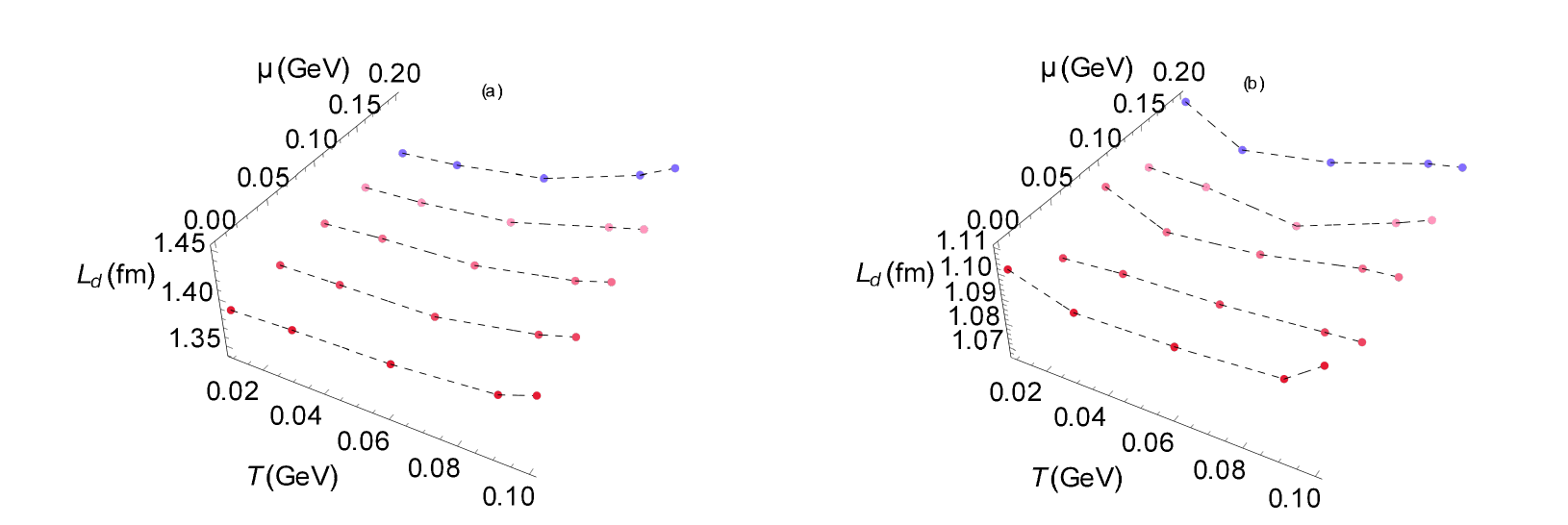}
    \caption{\label{re} effect of temperature/chemical potential on the string-breaking distance. Panel (a): equilateral triangular geometry. Panel (b): symmetric collinear geometry.}
    \label{fig:re}
\end{figure}

We plotted the 3D landscapes showing the string-breaking distances for QQQ(A) and QQQ(B) varying with temperature and chemical potential.

\begin{figure}[H]
    \centering
	\includegraphics[width=12cm]{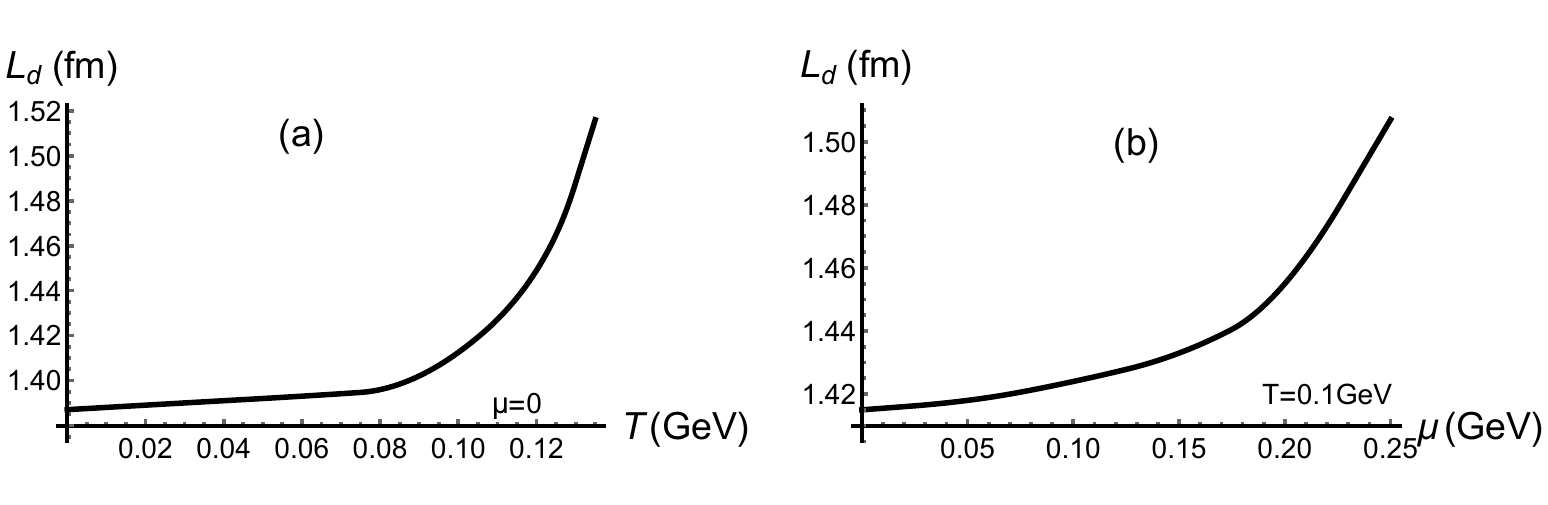}
    \caption{\label{ds}The string-breaking distance of the QQQ(A) configuration as a function of temperature and chemical potential is shown in the graph. Panel (a) represents the string-breaking distance at different temperatures when the chemical potential $\mu = 0$. Panel (b) represents the string-breaking distance at different chemical potentials when the temperature $T = 0.1$  GeV.}
    \label{fig:ds}
\end{figure}

\begin{figure}[H]
	\centering
    \includegraphics[width=12cm]{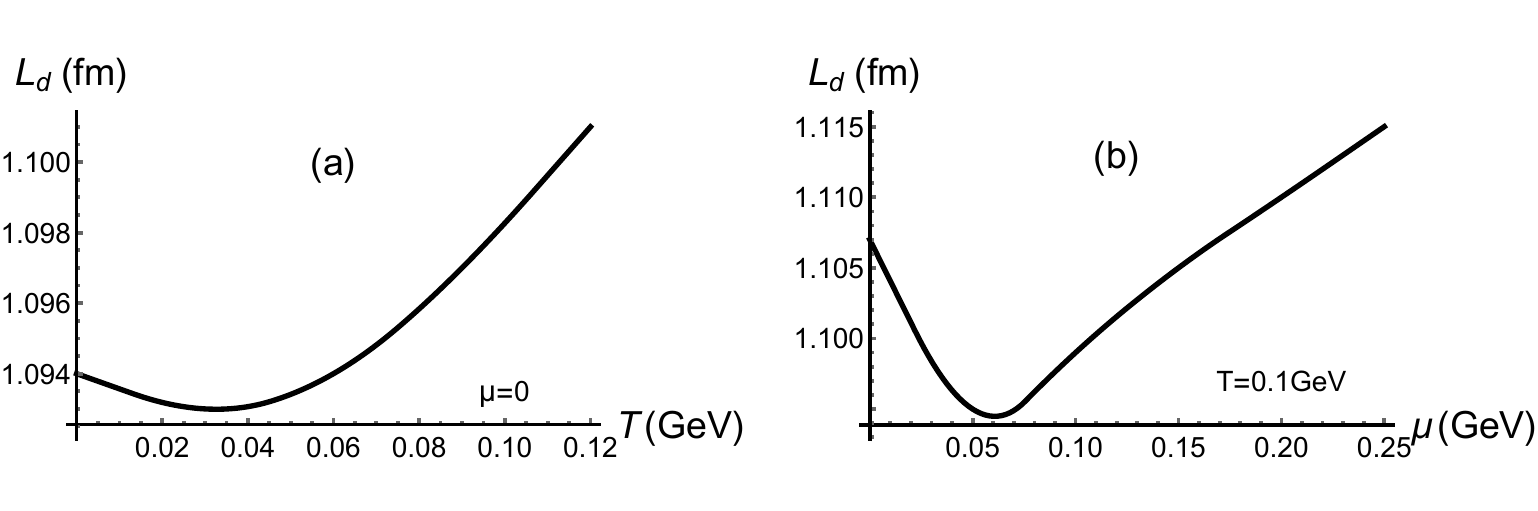}
    \caption{\label{gs}The string-breaking distance of the QQQ(B) configuration as a function of temperature and chemical potential is shown in the graph. Panel (a) represents the string-breaking distance at different temperatures when the chemical potential $\mu = 0$. Panel (b) represents the string-breaking distance at different chemical potentials when the temperature $T = 0.1$  GeV.}
    \label{fig:gs}
\end{figure}

The 3D relationships are depicted in Fig.~\ref{re}. From the 3D landscapes, we extracted 2D distance-temperature and distance-chemical potential plots, as presented in Fig.~\ref{ds} and Fig.~\ref{gs}, to investigate the effect of each thermodynamic factor individually.

It can be observed from the 2D plots that the effect of temperature and chemical potential on the string-breaking distance is more homogeneous in the equilateral triangular geometry while more complex in the symmetric collinear geometry. This phenomenon indicates the significant role of geometric configurations in determining the response of string-breaking distances to thermodynamic variations, which provides valuable insights.

For QQQ(A) equilateral triangular geometry, the string-breaking distance exhibited a monotonic increasing trend as the temperature and chemical potential arose. This behavior indicates that the unique triangular configuration contributes to the consistent increment of string-breaking distance.

In contrast, for QQQ(B) symmetric dyadic geometry, the string-breaking distance exhibits a non-monotonic trend, decreasing and then sharply increasing before exceeding a threshold. This nonlinear behavior implies a complex transformation of the mechanical equilibrium conditions associated with the dyadic configuration. The underlying mechanisms need to be further investigated.

Importantly, chemical potential has a more pronounced effect on changing the string-breaking distances than temperature. This underscores the vital role of chemical potential in the dynamics and hadronization of triply heavy quarks.

\section{Summary and conclusion}\label{Summary and conclusion}
The deformed AdS-RN background model discussed in this work effectively explores the string-breaking processes of triply heavy quarks. We systematically investigated the dependence of string-breaking distances on two geometric configurations of triply heavy quarks (equilateral triangular and symmetric collinear), finding that the differences in configurations significantly impact the trends in distance changes. We discovered that chemical potential has a more pronounced influence on changes in string-breaking distances compared to temperature. This study provides valuable new insights into the string-breaking distances of triply heavy quarks under varied temperature and chemical potential conditions.

\section{Acknowledgments}
This work is supported by the Research Foundation of Education Bureau of Hunan Province, China(Grant No. 21B0402 and No. 20C1594) and the Natural Science Foundation of Hunan Province of China under Grants No.2022JJ40344, and National Natural Science Foundation of China (NSFC) under the Grant No 12350410371 (M. A. Martin Contreras).  
\vspace{10mm}

\vspace{-1mm}
\centerline{\rule{80mm}{0.1pt}}

\end{document}